\documentclass[
 aip,
 apl,
 amsmath,amssymb,
 reprint,
]{revtex4-1}

\usepackage{graphicx}
\usepackage{dcolumn}
\usepackage{bm}
\usepackage{color}

\begin{document}
\preprint{AIP/123-QED}

\title[]{Probing the magnetic moment of FePt micromagnets prepared by Focused Ion Beam milling}

\author{H.C. Overweg}
\author{A.M.J. den Haan}
\author{H.J. Eerkens}

 \affiliation{Huygens-Kamerlingh Onnes Laboratory, Leiden University, Niels Bohrweg 2, 2333CA Leiden, the Netherlands}
\author{P.F.A. Alkemade }
 \affiliation{Kavli Institute of Nanoscience, Delft University of Technology, Lorentzweg 1,
2628 CJ Delft, The Netherlands}

\author{A.L. La Rooij }
\author{R.J.C. Spreeuw}
 \affiliation{Van der Waals-Zeeman Institute, University of Amsterdam,
Science Park 904, 1090 GL Amsterdam, The Netherlands}
\author{L. Bossoni}
\author{T.H. Oosterkamp}
 \email{oosterkamp@physics.leidenuniv.nl}
 \affiliation{Huygens-Kamerlingh Onnes Laboratory, Leiden University, Niels Bohrweg 2, 2333CA Leiden, the Netherlands}

\date{\today}

\begin{abstract}
We investigate the degradation of the magnetic moment of a 300 nm thick FePt film induced by Focused Ion Beam (FIB) milling. A  $1~\mu \mathrm{m} \times 8~\mu \mathrm{m}$ rod is milled out of a film by a FIB process and is attached to a cantilever by electron beam induced deposition. Its magnetic moment is determined by frequency-shift cantilever magnetometry. We find that the magnetic moment of the rod is $\mu = 1.1 \pm 0.1 \times 10 ^{-12} \mathrm{Am}^2$, which implies that 70 \% of the magnetic moment is preserved during the FIB milling process. This result has important implications for atom trapping and magnetic resonance force microscopy (MRFM), that are addressed in this paper.
\end{abstract}

\maketitle

The fabrication and characterization of micron sized permanent magnets is necessary for a broad range of applications, such as magnetic tweezers,\cite{Crick1950,Smith1992} magnetic imaging,\cite{Sidles1991,Marohn2012} and atom trapping with chips.\cite{Sidorov2001}

These chips are planar structures that generate magnetic fields, which are widely used to control ultra-cold atoms.\cite{Whitlock2009a} The incorporation of permanent magnets in atom chips offers several advantages over the use of current carrying wires:\cite{Sidorov2001, Davis1999} they dissipate no heat and allow more complex trap shapes. Moreover, permanent magnets can create larger field gradients, which facilitates tighter confinement of atoms,\cite{Xing2004} resulting in shorter time scales in trapping experiments. This does require the magnets to be patterned on small length scales. One of the materials currently under investigation is FePt in its L1$_0$ phase, a corrosion resistant material with high magnetocrystalline anisotropy.\cite{Xing2004,Xing2007,Gerritsma2007a} FePt atom traps that are currently in use are made by optical lithography and plasma etching.\cite{Gerritsma2007a,Leung2014a} The currently used patterns have length scales on the order of 10 $\mu$m.\cite{Jose2014a}

Micron sized magnets can also be used as a field gradient source for magnetic resonance force microscopy (MRFM).\cite{Sidles1991} This is a technique that uses a small magnet mounted on an ultrasoft cantilever to measure the magnetic interaction with spins in a sample underneath the cantilever. It thereby combines the advantage of elemental specificity of conventional Magnetic Resonance Imaging (MRI) techniques with the local and very sensitive probing techniques of Atomic Force Microscopy (AFM).\cite{Degen2009,*Poggio2010} Required properties for MRFM magnets are high magnetocrystalline anisotropy and a large remanent field.\cite{Stipe2001a} Small dimensions of the magnet are beneficial too, as they result in large magnetic field gradients, which increase the sensitivity of measurements.\cite{Jenkins2004,*Poggio2007} These requirements are similar to the requirements for atom traps and are all fulfilled by the aforementioned FePt.

One of the techniques to pattern FePt films is to use a Focused Ion Beam (FIB). However, FIB milling can damage the film, possibly degrading the magnetic properties. Examples of such damage include implantation of ions and other ion beam induced alterations to the crystal structure.\cite{Rubanov2004,*Giannuzzi1999} Determining the magnetic moment after FIB exposure is crucial for applications in both atom trapping and MRFM experiments.

In this letter, the damage caused by FIB milling on an FePt film is quantified by measuring the magnetic moment of a micron sized rod, which has been milled out of the film, and comparing it to the expected magnetic moment calculated from its volume and its remanent field. The rod is attached to a cantilever and its magnetic moment is determined by cantilever magnetometry, a sensitive technique to determine small magnetic moments.\cite{Stipe2001a,*Rossel1996} We demonstrate that FIB milling is a suitable way to shape magnetic films for atom trapping experiments and to prepare probes for MRFM.


The $300 \pm 10 \mathrm{nm}$ thick FePt film has been made at the Almaden Research Center of Hitachi.  Films of FePt have been sputtered on a Si substrate with a thin RuAl underlayer and a Pt interlayer at a temperature of $400^{\circ} \mathrm{C}$. This growth process leads to FePt in its L1$_0$ phase, which has a particularly high out-of-plane magnetization.  \cite{Shen2005}

\begin{figure}
\includegraphics[width=8.5 cm]{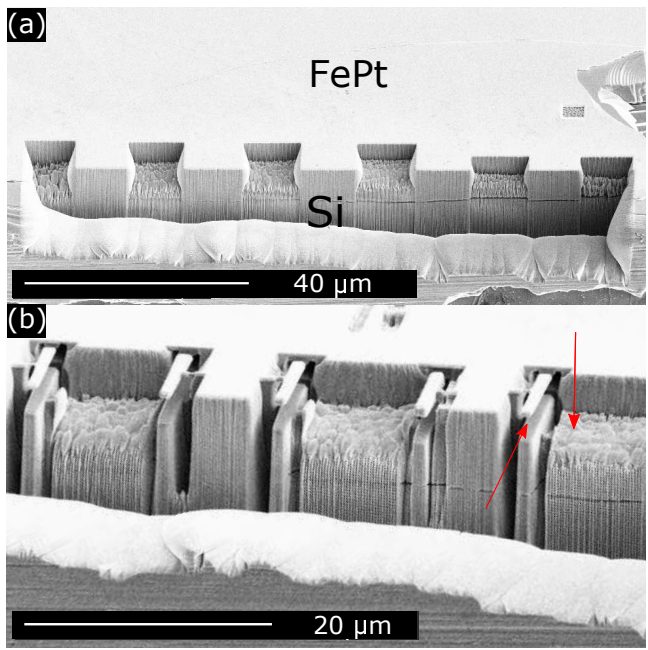}
\caption{\label{fig:fib} Fabrication of rods at the edge of an FePt film sputtered on a Si wafer: (a) crenelation of the edge (b) five rods at the end of the FIB process. The material has been milled from two perpendicular directions, see arrows.}
\end{figure}

As a first step to create rods, an indentation in the edge of the film is made with a FIB (Ga$^+$-ions, 30 keV,  7 nA ion current, Strata 235 Dual Beam from FEI). The edge is then crenelated (Fig. \ref{fig:fib}(a)) (ion current 500 pA) and rods are created in the sides of the crenels (Fig. \ref{fig:fib}(b)). The dimension of the rods is 8.1 $\mu$m in length, 1 $\mu$m in width and 1 $\mu$m in height (consisting of 300 nm FePt and 700 nm substrate). The sample is rotated by 90$^\circ$ to remove the material underneath the rods. The geometry facilitates the access necessary to mount a rod onto a cantilever.

\begin{figure}
\includegraphics[width = 8.5cm]{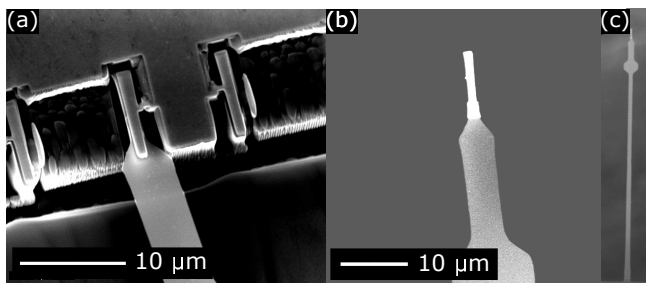}
\caption{\label{fig:sem} Fixation of a rod to a cantilever: (a) the cantilever is brought in position using a nanomanipulator. After an electron beam induced deposition (EBID) process to fix the rod to the cantilever, the connection to the film is broken by retracting the cantilever (b).  The widening on the cantilever works as a mirror for laser interferometry. (c) the cantilever-magnet assembly.}
\end{figure}

The FePt film and a cantilever (a single-crystalline silicon beam \cite{Chui2003}) are then placed on two stages of an in-house developed nanomanipulator \cite{Heeres2010} inside a Scanning Electron Microscope (NanoSEM 200 from FEI, USA). Using the nanomanipulator, we bring the cantilever in contact with an FePt rod (Fig. \ref{fig:sem}(a)). Subsequently, fixation is achieved by an electron beam induced deposition process with Pt(PF$_3$)$_4$ as a precursor gas. The last connection between the rod and the film is broken by suddenly retracting the cantilever. The finished assembly of the cantilever and the rod is shown in Fig. \ref{fig:sem}(b) and \ref{fig:sem}(c).

\begin{figure}[!ht]
\includegraphics[width= 8.5 cm]{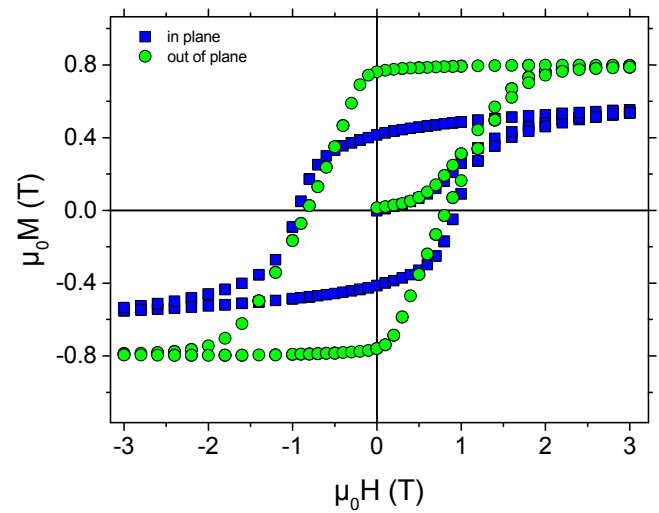}
\caption{\label{fig:hys} Magnetization of the film as a function of external magnetic field strength for two different orientations of the sample. For the out-of-plane
orientation the remanent field $\mu_0 M = 0.76 \pm 0.03~\mathrm{T}$ and for the in-plane orientation it is $\mu_0 M = 0.50 \pm 0.03~\mathrm{T}$. }
\end{figure}

Prior to the fabrication of the rods, the magnetization loop has been measured for a film of size 3 mm x 3 mm x 300 nm in a SQUID magnetometer (Quantum Design MPMS-5S). The measurement has been performed at room tempeature in two different geometries (Fig. \ref{fig:hys}): with an in-plane and an out-of-plane external field \textbf{H}. The remanent magnetization is $\mu_0 M = 0.76 \pm 0.03~\mathrm{T}$ for the out-of-plane geometry, while it is $\mu_0 M = 0.50 \pm 0.03~\mathrm{T}$ for the in-plane geometry. In Fig. \ref{fig:hys}, the remanant magnetic moment shows negligible dependence on the external magnetic field. This is expected for FePt, as the coercivity increases when the lateral size decreases.\cite{Attane2011} Therefore, the external field used in the cantilever magnetometry experiment should not affect the magnetic moment of the rod.

The rods are magnetized in a 3 T field at room temperature along the out-of-plane direction (i.e. along the direction of motion of the cantilever), to achieve a higher remanent field.

\begin{figure}
\includegraphics[width=8.5 cm]{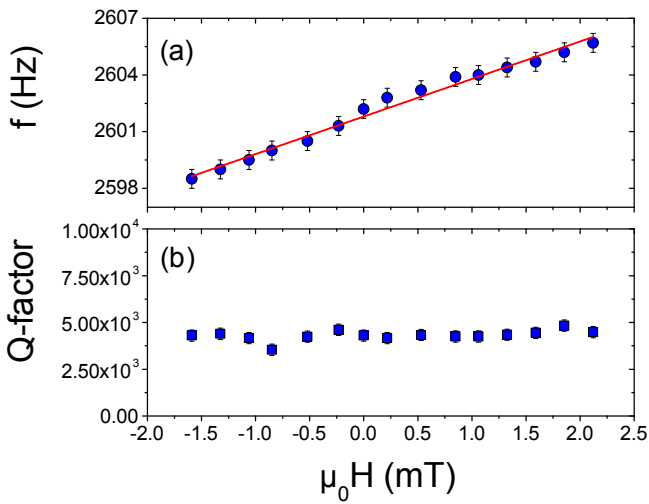}
\caption{\label{fig:mag} (a) Resonance frequency as a function of the external magnetic field determined from the cantilever's thermal spectrum. The slope of the curve implies a magnetic moment of $1.1 \pm0.1 \times 10^{-12}~\mathrm{Am}^2$, which means a volume of $0.8 \pm0.1~\mu \mathrm{m}^3$ has been damaged by the FIB. (b) The quality factor as a function of the external magnetic field as determined by a ring-down measurement.}
\end{figure}

Subsequently, dynamic-mode cantilever magnetometry is performed at room temperature at a pressure of 10$^{-5}$ mbar. The external magnetic field is provided by a Helmholtz coil of approximately 300 turns, generating magnetic fields up to 2 mT. The external magnetic field points along the direction of motion of the cantilever. To determine the magnetic moment $\mu$ of the rod, the resonance frequency is measured as a function of magnetic field strength. 
A fiber optic interferometer working at a wavelength of 1550 nm is used to detect the cantilever motion. The resonance frequency is determined by fitting the thermal motion of the cantilever's fundamental mode to a Lorentzian curve. A ring-down measurement, shown in Fig. \ref{fig:mag}(b), provides a more accurate measure of the quality factor Q.

The resonance frequency as a function of magnetic field is shown in Fig. \ref{fig:mag}(a). For the low magnetic field regime, the frequency shift $\Delta f$ as a function of magnetic field $\textbf{H}$ is given by:\cite{Marohn1998}
\begin{equation}
\label{eq}
\Delta f = \frac{f_0}{2k}\left(\frac{\alpha}{l}\right)^2  \mu \mu_0 H
\end{equation}

where $f_0$ is the resonance frequency in the absence of a magnetic field, $l = 200~\mu\mathrm{m}$ is the length of the cantilever, $\alpha = 1.377$ is a constant factor derived for beam cantilevers, and $k =3.3~\pm~0.2 \times 10^{-5}~\mathrm{N/m}$ is the stiffness of the cantilever, determined by the ‘added-mass method’. \cite{Cleveland1993}

Making use of equation \ref{eq}, the magnetic moment of the cantilever is deduced to be $\mu = 1.1 \pm0.1 \times 10^{-12}~ \mathrm{Am}^2$. Given the remanent magnetization of the FePt film and the volume of the magnet of $(1.00~\pm~0.02) ~\mu\mathrm{m} \times (8.10~\pm~0.02) ~\mu\mathrm{m} \times (0.30~\pm~0.01) ~\mu\mathrm{m}$, we would have expected a magnetic moment of $\mu = 1.5~\pm~0.1 \times 10^{-12}~ \mathrm{Am}^2$, if the magnet had been unaffected by the FIB process. The comparison shows that roughly 60 to 80 \% of the magnetic moment is preserved during the FIB process.  As both SQUID magnetometry and cantilever magnetometry allow only for the determination of the overall magnetic moment, we cannot precisely determine the damage profile.

The quality factor seems not to depend on the magnetic field strength. Ng et al.\cite{Jenkins2006} did report on a decrease of the quality factor in a magnetic field ranging up to 6 T. This change is negligible in the 2 mT magnetic field range we studied.

More FePt magnets have been attached to cantilevers by the procedure described above. However, the orientation of the out-of-plane direction of the FePt film with respect to the direction of motion of these cantilevers was different (see supplemental material\cite{sub} for more information). Though beneficial for MRFM experiments,\cite{Marohn1998} these probes are unfit for cantilever magnetometry experiments.

We believe MRFM would benefit from the described force sensor. Since the force exerted by a spin in the sample on the cantilever is proportional to the gradient of the magnetic field, it is beneficial to use small magnets.
In our previous work, we employed NdFeB spheres with a diameter of 3 $\mu$m. \cite{Vinante2012} The field gradient cannot be increased by using smaller NdFeB particles, because they seem to lose their magnetization when scaled down further.  \cite{Hammel}  Even though FePt has a remanant magnetization which is roughly half as large as that of NdFeB, the possibility to create smaller magnets is promising for the sensitivity of MRFM experiments.
The larger magnetic field gradient is not the only improvement that small FePt magnets would yield. It has been observed that the quality factor of MRFM cantilevers can drop drastically when approaching the sample surface.\cite{Vinante2011} This is most likely due to a dissipative interaction of spins in the sample with the magnet. A smaller magnet interacts with fewer spins and therefore suffers less from this unwanted damping.
A forthcoming experiment will enable us to quantify the improvement in the resolution provided by the FePt rods.

Concerning atom trapping, the factor limiting the resolution of FePt traps created by optical lithography and plasma etching is the redeposition of the etched material, the magnetic properties of which are unknown.\cite{Gerritsma2007} SEM images show that this redeposition can be of the order of several hundreds of nanometers.
From SEM images made after FIB milling, we conclude that for the FePt rods described in this paper redeposition of FePt is negligble compared to the loss of magnetic volume caused by the FIB milling process. Furthermore, the damage induced can possibly be reduced by using a helium FIB. Hence FIB milled patterns could have an advantage over patterns created by optical lithography and plasma etching, when aiming for trap sizes on the order of a micrometer.\cite{Leung2011,Singh2014} For the formation of such traps a better understanding of the shape of the damaged region of magnetic films would be needed.
FIB milling of FePt will probably not suffice to go to an atom trap scale of the order of 100 nm. Electron beam lithography is the most suitable technique when aiming for submicrometer sizes. \cite{Leung2011} This method is currently used in various groups.

We have shown a fabrication process for micrometer size FePt magnets  by FIB milling and a way to attach these magnets to ultrasoft cantilevers by electron beam induced deposition. This technique could in principle be used for any magnetic film. From cantilever magnetometry measurements we conclude that 60 to 80 \% of the magnetic moment is preserved during the FIB milling process. FIB milled magnets could therefore be used in atomic trapping experiments when aiming for a trap size on the order of a micrometer. The magnet attached to the cantilever can be used as a probe in MRFM experiments. The small dimensions of the magnet are expected to improve the sensitivity of MRFM.

\begin{acknowledgments}
The authors thank J.J.T. Wagenaar for fruitful discussions.
This work was supported in part by Fundamenteel Onderzoek der Materie (FOM) and by the Netherlands organization of scientific research (NWO).
\end{acknowledgments}

\bibliographystyle{aipnum4-1}

\end{document}